\documentstyle[pre,aps]{revtex}

\newcommand{\br}{{\bf r}}
\newcommand{\be}{\begin{equation}}
\newcommand{\ee}{\end{equation}}
\newcommand{\dgr}{\dagger}
\newcommand{\dlt}{\delta}
\newcommand{\prt}{\partial}
\newcommand{\vp}{\varphi}
\newcommand{\om}{\omega}
\newcommand{\bt}{\beta}
\newcommand{\al}{\alpha}
\newcommand{\Dlt}{\Delta}
\newcommand{\ra}{\rightarrow}
\newcommand{\gm}{\gamma}
\newcommand{\ep}{\varepsilon}
\newcommand{\Om}{\Omega}
\newcommand{\Gm}{\Gamma}

\begin{document}

\draft

\title{Nonlinear Coherent Modes of Trapped Bose-Einstein
Condensates}
\author{V.I. Yukalov$^{1,2}$, E.P. Yukalova$^{1,3}$, and
V.S. Bagnato$^1$}
\address{$^1$Instituto de Fisica de S\~ao Carlos, Universidade de S\~ao Paulo\\
Caixa Postal 369, S\~ao Carlos, S\~ao Paulo 13560-970, Brazil}

\address{$^2$Bogolubov Laboratory of Theoretical Physics \\
Joint Institute for Nuclear Research, Dubna 141980, Russia}

\address{$^3$Department of Computational Physics, Laboratory of
Information Technologies \\
Joint Institute for Nuclear Research, Dubna 141980, Russia}

\maketitle

\begin{abstract}

Nonlinear coherent modes are the collective states of trapped Bose
atoms, corresponding to different energy levels. These modes can
be created starting from the ground state condensate that
can be excited by means of a resonant alternating field. A thorough
theory for the resonant excitation of the coherent modes is presented.
The necessary and sufficient conditions for the feasibility of this
process are found. Temporal behaviour of fractional populations and
of relative phases exhibits dynamic critical phenomena on a critical
line of the parametric manifold. The origin of these critical phenomena
is elucidated by analyzing the structure of the phase space. An atomic
cloud, containing the coherent modes, possesses several interesting
features, such as interference patterns, interference current, spin
squeezing, and massive entanglement. The developed theory suggests
a generalization of resonant effects in optics to nonlinear systems
of Bose-condensed atoms.

\end{abstract}


\newpage

\section{Introduction}

Low-temperature properties of dilute Bose-Einstein condensate are
well described by the Gross-Pitaevskii equation (see reviews [1--3]).
The mathematical structure of the latter equation is that of the
nonlinear Schr\"odinger equation. This equation in the presence of
a confining potential possesses a discrete spectrum of stationary
states. The ground-state solution corresponds to the equilibrium Bose
condensate. If it could be possible to macroscopically populate a
nonground-state energy level, this would correspond to the creation
of a nonground-state Bose-Einstein condensate. The feasibility of
creating such nonground-state condensates was advanced in Ref. [4].
The properties of these excited self-consistent states and different
ways of their formation have been previously discussed in [4--12].
A nonlinear dipole mode in a two-component condensate was excited
in experiment [13].

A few words are in order to clarify the meaning of the self-consistent
excited states. First of all, they should not be confused with the
elementary collective excitations. The latter correspond to small
oscillations around a given trap state, say around the ground state,
and are described by the {\it linear} Bogolubov-De Gennes equations.
While the self-consistent trap states are described by the
{\it nonlinear} Gross-Pitaevskii equation. This is why the stationary
solutions to the latter equation are called the {\it nonlinear modes}.
The Bose-condensed atoms possess a high degree of coherence [1--3],
and the related nonlinear Schr\"odinger equation is known to describe
a coherent wave function, or the coherent field [14]. Because of this,
the nonlinear trap modes are termed the {\it coherent modes}. The
atomic densities, corresponding to different nonlinear coherent modes,
possess rather different spatial distribution, because of which such
modes are sometimes named as topological [3,13]. Vortices are an example
of these modes. The time-dependent nonlinear Schr\"odinger  equation
supports the existence of solutions called solitons [15,16]. The latter
are certainly nonlinear by their origin, but they are not necessarily
identical to the nonlinear coherent modes. Only those solitons can be
identified with such modes, which are the solutions to the stationary
eigenvalue problem.

In the present paper, we study the excitation of the nonlinear
coherent modes by means of the resonance pumping technique suggested
in Ref. [4]. To some extent, this is analogous to the problem of
periodically modulated nonlinear optical fibers [17], as opposed to
the adiabatic perturbations of nonlinear soliton equations [18]. Our
aim here is to develop a thorough theory of the nonlinear coherent
modes, analyzing their features that have not been considered in our
previous publications [4,7]. The main new points are as follows:

\vskip 2mm

(1) A more general pumping field is considered, which makes it easier to
create vortices and, what is more important, allows the variation of the
initial phase difference related to the resonantly connected fractional
populations. Varying this phase difference makes the dynamical picture
much richer, providing several novel regimes of motion.

\vskip 2mm

(2) Necessary and sufficient conditions for the feasibility of the
resonant excitation of the nonlinear modes are derived. These conditions
impose limitations on the admissible strength of atomic interactions
and on the number of trapped atoms. The found restriction on the number
of atoms is similar to the critical number arising in the analysis of
the stability of condensates with a negative scattering length. The ways
of optimizing the found limits are discussed.

\vskip 2mm

(3) Temporal dynamics of fractional populations and of the related
phase difference is studied in detail for all admissible initial
conditions and for varying system parameters. Stability analysis is
accomplished. Investigating the phase portrait determines possible
dynamic phase transitions.

\vskip 2mm

(4) A special attention is paid to dynamic critical phenomena,
discovered earlier [19,20] by means of numerical calculations, but
which have not received a comprehensive explanation. The origin of
these critical phenomena is elucidated as being due to the saddle
separatrix crossing by the starting point of a trajectory in phase
space. A set of critical lines on the parametric manifold is found.

\vskip 2mm

(5) The influence of power broadening on the long lasting resonant
pumping is considered. This effect, despite well preserved resonance
conditions on the excitation frequency, may lead to the accumulation
of perturbations, resulting in the appearance of attenuation, limiting
the admissible time of pumping. This limiting time is found to be of
order of the trapping lifetime caused by depolarizing atomic collisions.

\vskip 2mm

(6) Interference pattern and interference current are found to display
two-scale oscillations, slow and fast. These interference effects can be
experimentally observed.

\vskip 2mm

(7) Atomic spin squeezing in the presence of multiparticle entanglement
is studied. The dispersion of the population-difference operator is
always squeezed with respect to that of the transition-dipole operator.
This means that the fractional populations can be measured with a
precision that is higher than the accuracy of measuring transitional
characteristics, such as the relative phase. The connection of these
results with the notions of state coherence and transition coherence
is discussed.

\section{Coherent Modes}

The Gross-Pitaevskii equation is usually treated as an approximate
mean-field equation for the broken symmetry order parameter [1,2]. This
interpretation, however, is not unique. Here we show that this equation
can be considered as an {\it exact} equation for a coherent wave
function. The latter interpretation sheds some light on the intimate
relation between Bose-Einstein condensate and coherence. At the same
time, this allows us to give a rigorous mathematical definition of
coherent modes which will be considered in the following sections. To
suggest an accurate definition of the nonlinear coherent modes is of
principal importance, since these are the basic object studied in the
present paper.

In the field approach, dealing with the field operators $\psi(\br,t)$,
the standard multiatomic Hamiltonian reads
$$
H = \int \psi^\dgr (\br,t)\left [ -\; \frac{\hbar^2{\bf\nabla}^2}{2m_0}
+ U(\br,t)\right ] \psi(\br,t)\; d\br +
$$
\be
\label{1}
+ \frac{1}{2} \int \psi^\dgr(\br,t)\; \psi^\dgr(\br',t)\;
\Phi(\br -\br') \; \psi(\br',t)\; \psi(\br,t) \; d\br\; d\br' \; ,
\ee
where $m_0$ is atomic mass; $U(\br,t)$, external potential; and
$\Phi(\br)$, interaction potential. The field operators satisfy the
Bose commutation relations. Trapped atoms constitute a dilute gas,
for which the interaction potential can be presented in the Fermi
form
\be
\label{2}
\Phi(\br) = A_s \dlt(\br) \; , \qquad A_s \equiv 4\pi\hbar^2\;
\frac{a_s}{m_0} \; ,
\ee
in which $a_s$ is an $s$-wave scattering length.

Coherent states [21] are defined as the eigenstates of the annihilation
operator
\be
\label{3}
\psi(\br,t)\; |\eta\rangle \; = \eta(\br,t)\; |\eta \rangle \; .
\ee
Here $|\eta\rangle$ is a {\it coherent state} and the function
$\eta(\br,t)$, playing the role of an eigenvalue, is called the coherent
wave function, or, for brevity, the {\it coherent field}. The Heisenberg
equation of motion for the field operator yields
\be
\label{4}
i\hbar\; \frac{\prt}{\prt t} \; \psi(\br,t) = H[\psi]\; \psi(\br,t) \; ,
\ee
with the nonlinear Hamiltonian
\be
\label{5}
H[\psi] \equiv  - \; \frac{\hbar^2{\bf\nabla}^2}{2m_0} +
U(\br,t) + A_s \psi^\dgr(\br,t) \; \psi(\br,t) \; .
\ee
Taking the average of Eq. (4) over the coherent state
$|\eta\rangle$ gives the equation
\be
\label{6}
i\hbar \; \frac{\prt}{\prt t}\; \eta(\br,t) = H[\eta]\; \eta(\br,t)
\ee
for the coherent field, with the nonlinear Hamiltonian
\be
\label{7}
H[\eta] \equiv - \; \frac{\hbar^2{\bf\nabla}^2}{2m_0} +
U(\br,t) + A_s|\eta(\br,t)|^2 \; .
\ee
At zero temperature, when all $N$ atoms are condensed into a coherent
state, the coherent field $\eta(\br,t)$ is normalized to $N$. It is
convenient to define
\be
\label{8}
\eta(\br,t) \equiv \sqrt{N}\; \vp(\br,t) \; .
\ee

The external potential
\be
\label{9}
U(\br,t) = U(\br) + V(\br,t)
\ee
consists of a stationary trapping potential $U(\bf r)$ and of an
additional perturbing potential $V(\br,t)$. Then Eq. (6) for the
coherent field can be written as
\be
\label{10}
i\hbar \; \frac{\prt}{\prt t} \; \vp (\br,t) = \left (
\hat H[\vp] + \hat V\right ) \vp(\br,t) \; ,
\ee
where $\hat V\equiv V(\br,t)$ and
\be
\label{11}
\hat H[\vp] \equiv - \; \frac{\hbar^2{\bf\nabla}^2}{2m_0} +
U(\br) + NA_s |\vp(\br,t)|^2 \; .
\ee

Equations (6) and (10) are exact equations for the coherent fields.
Such equations in physical literature are termed the time-dependent
Gross-Pitaevskii equations. In mathematical parlance, these are the
nonlinear Schr\"odinger equations [15,16]. Being complimented by
initial and boundary conditions, such equations compose the Cauchy
problem. For nonlinear differential equations in partial derivatives,
the Cauchy problem possesses a unique analytic solution under the
conditions of the Cauchy-Kovalevskaya theorem [22]. These conditions
are not fulfilled for Eqs. (6) or (10). Hence these equations may,
in general, have several solutions. To guarantee the uniqueness of
solution, it is necessary to impose some additional constraints
motivated by the physics of the problem [15,16]. In particular, it is
necessary to specify a functional space to which the sought solution
is supposed to pertain. For the studied equations, it is possible to
proceed as follows.

Let us consider the eigenproblem
\be
\label{12}
\hat H[\vp_n]\; \vp_n(\br) = E_n\; \vp_n(\br) \; ,
\ee
defining the stationary solutions $\vp_n(\br)$ labelled by a multi-index $n$. 
In the presence of a confining potential $U(\br)$, the spectrum $\{ E_n\}$ 
is discrete. The countability of the set $\{\vp_n(\br)\}$ for a nonlinear 
problem is not absolutely guaranteed. For the Gross-Pitaevskii equation with 
a single-well confining potential, the eigenproblem can be treated as an 
analytical continuation of that for the associated linear Schr\"odinger 
equation resulting in the limit of vanishing nonlinearity [10,12]. Then the 
set $\{\vp_n(\br)\}$ is easily countable. But in general, e.g. in the case 
of multiwell confining potentials, there appear the stationary solutions 
to the Gross-Pitaevskii equation, which have no linear counterparts [12], 
because of which the countability problem becomes rather nontrivial. However, 
for what follows, the countability of the set of stationary solutions to 
Eq. (12) is not compulsory. The main is that these solutions be appropriately 
labelled by a multi-index. The solutions $\vp_n(\br)$ to the nonlinear 
eigenproblem (12) are the {\it nonlinear coherent modes}. Generally, some 
of the levels can be degenerate, which does not hamper the consideration.
Moreover, the problem of degeneracy, when it becomes annoying, can be avoided
by adding to the Hamiltonian $\hat H[\vp_n]$ terms lifting the degeneracy.
Such auxilliary terms may be removed after accomplishing the required
procedures. The main is that the set $\{\vp_n(\br)\}$ be composed of all
possible different solutions. This, however, does not presuppose that all
functions from the set are mutually orthogonal, so that the scalar product
\be
\label{13}
(\vp_m,\vp_n) \equiv \int \vp_m^*(\br)\; \vp_n(\br)\; d\br
\ee
is not compulsory a Kroneker delta. The possible nonorthogonality of the
coherent modes may be due to the fact that the Hamiltonian $\hat H[\vp]$
is nonlinear. In such a case, one has
$$
\left ( \hat H[\vp_m]\vp_m,\; \vp_n\right ) - \left ( \vp_m,\;
\hat H[\vp_n]\vp_n\right ) = (E_m - E_n)(\vp_m,\vp_n) \; .
$$
The left-hand side of the latter equality, for the nonlinear Hamiltonian
(11), reads
$$
\left ( \hat H[\vp_m]\vp_m,\; \vp_n\right ) - \left ( \vp_m,\;
\hat H[\vp_n]\vp_n\right ) = NA_s \left ( \vp_m,\;[|\vp_m|^2-|\vp_n|^2]
\vp_n \right ) \; ,
$$
which, in general, is not zero. Therefore, the modes $\vp_m$ and $\vp_n$,
corresponding to $E_m\neq E_n$, are not necessarily orthogonal. But the modes
can, of course, be always normalized, so that $(\vp_n,\vp_n)=1$. Now, let us
organize a closed linear envelope over the  set $\{\vp_n(\br)\}$, which, being
equipped with the scalar product (13), composes a Hilbert space
\be
\label{14}
{\cal H} \equiv \overline{\cal L}\{ \vp_n(\br)\} \; .
\ee
By this construction, the set $\{\vp_n(\br)\}$ is total [21], which
implies that a function from the Hilbert space (14) can be presented
as an expansion over this set. For instance, looking for a solution
of Eq. (10), pertaining to the space (14), we may write the mode
presentation
\be
\label{15}
\vp(\br,t) = \sum_n b_n(t) \vp_n(\br) \; .
\ee

One should not confuse the property of totality with that of
completeness. The totality of a set $\{ \vp_n(\br)\}$ means the
possibility of presenting a function $\vp\in{\cal H}$ from the Hilbert
space (14) as the sum (15). While the completeness of a set
$\{\vp_n(\br)\}$ implies the existence of the resolution of unity
$$
\sum_n \rho_n\; \vp_n(\br)\; \vp_n^*(\br') = \dlt(\br-\br') \; ,
$$
in general, involving a weighting set $\{\rho_n\}$. More details on
the delicate difference between totality and completeness can be found
in the book [21].

The property of totality is weaker than that of completeness. A
complete basis is certainly total, while a total set is not compulsory
complete. In our case, the set $\{\vp_n(\br)\}$ of coherent modes,
defined by the eigenproblem (12), may happen to be complete. A hint
to this is the observation [10] that the eigenstates of the nonlinear
problem (12) can be considered as an analytical continuation, with
respect to the nonlinearity constant $A_s$, of the eigenstates for
the related linear problem, which possesses a complete basis. Moreover,
the completeness of the countable set of eigenfunctions to the
one-dimensional nonlinear Schr\"odinger Hamiltonian has been rigorously
proved [23,24]. There is no such an exact proof of completeness for the
three-dimensional case. But we would like to stress that nowhere in
what follows we shall use the completeness of the set $\{\vp_n(\br)\}$
and will never invoke the related resolution of unity. The sole thing
we shall employ is the totality of the set of coherent modes, which is
equivalent to the presentation (15). Note that for a homogeneous gas,
the modes $\vp_n(\br)$ reduce to plane waves [25].

\section{Resonant Excitation}

In the case of a discrete spectrum, the energy levels are separated,
and the interlevel distance is characterized by the transition frequency
\be
\label{16}
\om_{mn} \equiv \frac{1}{\hbar}\left ( E_m - E_n\right ) \; .
\ee
In order to single out a pair of levels, it is necessary to connect them
by a resonant transition. For this, an external oscillatory field
\be
\label{17}
V(\br,t) = V_1(\br)\cos\om t + V_2(\br)\sin\om t
\ee
has to be imposed on the atomic system, with a frequency tuned to the
transition frequency between a couple of chosen levels, say the levels
with the energies $E_1<E_2$, so that the transition frequency be
\be
\label{18}
\om_{21} \equiv \frac{1}{\hbar} \left ( E_2 - E_1\right ) \; .
\ee
Then the resonance condition reads
\be
\label{19}
\left | \frac{\Dlt\om}{\om} \right | \ll 1 \qquad
(\Dlt\om \equiv \om -\om_{21} ) \; .
\ee

The resonant excitation of nonlinear modes in Bose-Einstein condensates is
rather analogous to resonant two-level transitions in isolated atoms [26].
The difference is that in atoms, one considers single-electron levels, while
in trapped Bose condensate, one deals with collective coherent states of many
atoms. To separate out a pair of levels, coupled by a resonant transition,
it is necessary that the spectrum be nonequidistant. This is the case for
electronic spectra in atoms. And, fortunately, this is also true for the
spectrum of coherent modes of trapped Bose condensates, where the spectrum
becomes nonequidistant due to atomic interactions [3,4]. It may happen,
that the transition frequencies (16), though being different for $E_m\neq
E_n$, are densely packed, as e.g. is the case for highly excited states
of an optical atom [26]. Then the resonance condition (19) has to be
strengthened, so that the detuning $\Dlt\om$ be smaller than the transition
frequencies between the energy levels neighboring the selected levels
with the energies $E_1$ and $E_2$. Such a situation, because of the
uncertainty relation $\Dlt\om\cdot\Dlt t\sim 1$, would lead to a long time
$\Dlt t$ required to excite a specific mode from a densely packed spectrum.
The same difficulty may arise for Bose-condensed atomic gases, although for
the latter one has more possibility of overcoming the problem, as compared
to the case of a single atom. This is because a single atom possesses a fixed
energy spectrum, while the spectrum of nonlinear coherent modes can be
modified by varying the number of particles $N$ or the strength $A_s$ of
atomic interactions, e.g., by means of Feshbach resonance. At the same time,
these interactions render the system nonlinear, which leads to some important
differences in the behaviour of Bose condensate, subject to the action of
a resonant field, as compared to optical resonance in atoms. Comparing these
two types of resonance, we may say that in a single atom, one can induce
a linear resonance, while in a collective of coherent atoms, there occurs
a {\it nonlinear coherent resonance}.

The behaviour of the coherent field (15) is described by the nonlinear
equation (10), where the alternating field (17) is in resonance with the
transition frequency (18) of a chosen interlevel transition. Earlier,
we have considered [4,7] only the transitions from the ground state to
an excited collective state. But if one is able to transfer atoms from
the ground state to an excited level, then by applying one more
alternating field, with a different resonant frequency, it is possible
to transfer atoms from one excited level to another higher level.
Therefore, we consider the general situation of an arbitrary pair of
collective levels coupled by a resonant transition.

The coefficients $b_n(t)$ in the mode presentation (15) can be written
as
\be
\label{20}
b_n(t) = c_n(t)\exp\left ( -\; \frac{i}{\hbar} \; E_n\; t\right ) \; ,
\ee
with $E_n$ being the $n$-mode energy. In order that different modes
could be separated, it is necessary that the coefficient $c_n(t)$ be
a slow function of time, as compared to the fast oscillations of the
exponential $\exp(-\frac{i}{\hbar}E_nt)$. This implies the validity of
the {\it necessary condition for mode separation}
\be
\label{21}
\frac{\hbar}{E_n}\left | \frac{dc_n}{dt} \right | \ll 1 \; .
\ee
Inequality (21) reminds us the slowly-varying amplitude
approximation, so common in optics [26]. Under this condition, we
may try to get a nontrivial mode picture, following the way accepted
in optical resonance. For this purpose, we substitute the mode form
(15) into the evolution equation (10), multiply the latter by
$\vp_n^*(\br)\exp(\frac{i}{\hbar}E_nt)$, integrate it over $\br$,
and average over time, treating $c_n(t)$ as quasi-invariants. The
mathematical foundation of this procedure is based on the averaging
technique [27]. In this way, we meet two transition amplitudes, one
that is caused by atomic interactions
\be
\label{22}
\al_{mn}\equiv A_s\; \frac{N}{\hbar}
\int |\vp_m(\br)|^2 \left [ 2|\vp_n(\br)|^2 - |\vp_m(\br)|^2
\right ]\; d\br \; ,
\ee
and another due to the alternating field,
\be
\label{23}
\bt_{mn} \equiv \frac{1}{\hbar} \int
\vp_m^*(\br) \left [ V_1(\br) - i V_2(\br)\right ] \vp_n(\br)\;
d\br \; .
\ee
Recall that in optics one has only the transition amplitude of an
external resonant field defining a Rabi frequency. Here, because of
atomic interactions, there appears an additional amplitude (22). Note
that for different energy levels, the amplitude (22) may have different
signs, for a fixed sign of the interaction parameter $A_s$.

The condition for mode separation (21) is necessary, however not
sufficient. For employing the averaging technique [27], we need that
the transition amplitudes be smaller than the related transition
frequencies,
\be
\label{24}
\left | \frac{\al_{mn}}{\om_{mn}}\right | \ll 1 \; , \qquad
\left | \frac{\bt_{mn}}{\om_{mn}}\right | \ll 1 \; .
\ee
This mathematical requirement for the applicability of the averaging
technique has a transparent physical meaning $-$ all perturbations,
either produced by internal interactions or by external fields, must
be smaller than the corresponding transition frequencies, in order
that the latter be well defined as such. Again, the situation here
is completely analogous to optics, where one requires that the Rabi
frequencies be smaller than the transition frequency, in order to avoid
the power broadening spoiling the resonance picture. In our case, the
difference with optics is that we have, in addition to the external
transition amplitude (23), an internal amplitude (22). The latter
appears because of the interaction term in the nonlinear Hamiltonian
(11). This term depends on time through the function $\vp(\br,t)$.
Therefore, one could naively decide that as soon as the coherent field
(15) changes from one mode to another, the resonance condition would
be broken. However, this does not happen if the restrictions (24) are
hold true. Then, despite the variation of the interaction term, the
transition frequencies are always well defined, making it possible
to tune a well preserved resonance.

Restrictions (24) are stronger than inequality (21). The latter is only
necessary, while the former are necessary and sufficient for mode
separation. Mathematically, conditions (24) allow us, when employing
the averaging technique, to use the integral
$$
\lim_{T\ra\infty} \; \frac{1}{T} \int_0^T
\exp(i\om_{mn}t)\; dt = \dlt_{mn} \; .
$$
As a result, we come to the set of equations
\be
\label{25}
i\; \frac{dc_n}{dt} = \sum_{m(\neq n)} \al_{mn}|c_m|^2 c_n +
\frac{1}{2}\; \dlt_{n1}\bt_{12} c_2 e^{i\Dlt\om t} +
\frac{1}{2}\; \dlt_{n2} \bt_{12}^* c_1 e^{-i\Dlt\om t} \; ,
\ee
for the functions $c_n=c_n(t)$, satisfying the normalization condition
$$
\sum_n | c_n|^2  = 1 \; .
$$
From Eq. (25) it immediately follows that, if $n\neq 1,2$, then one has
\be
\label{26}
c_n(t) = c_n(0) \exp\left\{ -i \sum_{m(\neq n)} \al_{nm}
\int_0^t |c_m(t')|^2\; dt'\right \} \; .
\ee
Supposing that at the initial time no levels, except $n=1,2$, are
populated, that is $c_n(0)=0$ for $n\neq 1,2$, we get from Eq. (26)
that
\be
\label{27}
c_n(t) = 0 \qquad (n\neq 1,2; \; t\geq 0) \; .
\ee
Hence, the normalization condition is
\be
\label{28}
|c_1|^2 + | c_2|^2  = 1\; .
\ee
And the set of equations (25) reduces to two equations
\be
\label{29}
i\; \frac{dc_1}{dt} = \al_{12}|c_2|^2 c_1 +
\frac{1}{2}\bt_{12} c_2 e^{i\Dlt\om t} \; , \qquad
i\; \frac{dc_2}{dt} = \al_{21}|c_1|^2 c_2 +
\frac{1}{2}\bt_{12}^* c_1 e^{-i\Dlt\om t} \; .
\ee
This reduction is valid provided that the conditions
\be
\label{30}
\left | \frac{\al_{12}}{\om_{21}}\right | \ll 1 \; , \qquad
\left | \frac{\al_{21}}{\om_{21}}\right | \ll 1 \; , \qquad
\left | \frac{\bt_{12}}{\om_{21}}\right | \ll 1 \;
\ee
are satisfied.

It would be tempting to postulate from the very beginning that, under
the resonance condition (19), only two modes are involved in the process.
However such a simplified approach would leave doubts of the possibility
to reduce the consideration to an effective two-level system. Therefore,
we felt it necessary to present an accurate mathematical derivation of
the reduced equations (29). Moreover, in the process of this derivation,
we found conditions (30) defining the region of applicability for the
resonant two-level picture.

By acting on the trapped Bose gas with several resonant fields, we
could separate out not two but several coherent modes, reducing the
consideration to a finite-level system. The possibility of creating
from Bose condensate the resonant finite-level systems is of great
importance. This allows for a wide variety of different applications,
analogous to those for resonant finite-level atoms in optics [26,28].
Since the properties of resonant Bose-condensed systems are rather
different from those of single resonant atoms, we may expect the
appearance of novel effects, such as dynamic critical phenomena
[19,20]. In what follows, for brevity, the effect of resonant formation
of nonlinear coherent modes in Bose-condensed gas will be called
{\it coherent resonance}.

Note that a resonant finite-level system of coherent modes is
principally different from the two-mode models accepted for describing
Bose condensates in stationary double-well potentials [8,29--30]. Though
some equations in our case may resemble some expressions in the case
of the double-well potentials, this similarity is rather formal, being
due to the simple fact that all finite-level systems share some common
properties in their mathematical structure. However the physics of a
system, subject to the action of a resonant alternating field,
drastically differs from what happens in stationary double wells.

\section{Transition Amplitudes}

The set of inequalities (19) and (30) provides us with the necessary and
sufficient conditions for realizing the coherent resonance. The frequency
and the amplitude of the alternating external field (17) can always be
chosen in order to satisfy the resonance condition (19) and the last of the
inequalities (30). Hence the main concern is connected with the validity of
the first two inequalities (30) for the internal transition amplitudes (22)
due to atomic interactions. To understand better what is the restriction
imposed by these inequalities, let us consider a harmonic cylindrical trap
with the radial frequency $\om_r\equiv \om_x=\om_y$ and axial frequency
$\om_z$, denoting the aspect ratio
\be
\label{31}
\nu \equiv \frac{\om_z}{\om_r} \; .
\ee
Introduce the dimensionless cylindrical variables
\be
\label{32}
r \equiv \frac{\sqrt{r_x^2+r_y^2}}{l_r} \; , \qquad
z \equiv \frac{r_z}{l_r} \qquad
\left ( l_r \equiv \sqrt{\frac{\hbar}{m_0\om_r}} \right ) \; ,
\ee
measured in units of the oscillator length $l_r$. In the dimensionless
coupling parameter
\be
\label{33}
g \equiv 4\pi\; \frac{a_s}{l_r}\; N \; ,
\ee
$a_s$ is an $s$-wave scattering length and $N$ is the number of
particles. The nonlinear Hamiltonian, measured in units of $\om_r$,
is
\be
\label{34}
\hat H = -\; \frac{1}{2}{\bf\nabla}^2 + \frac{1}{2}\left (
r^2 + \nu^2 z^2 \right ) + g|\psi|^2 \; ,
\ee
where
$$
\psi(r,\vp,z) \equiv l_r^{3/2} \vp(\br)
$$
is a dimensionless wave function and
$$
\nabla^2 = \frac{\prt^2}{\prt r^2} + \frac{1}{r}\;
\frac{\prt}{\prt r} + \frac{1}{r^2}\; \frac{\prt^2}{\prt\vp^2} +
\frac{\prt^2}{\prt z^2} \; .
$$
In this notation, the eigenproblem (12) reads
\be
\label{35}
\hat H \psi_{nmj}(r,\vp,z) = E_{nmj}\psi_{nmj}(r,\vp,z) \; ,
\ee
where $n=0,1,2,\ldots$ is the radial quantum number,
$m=0,\pm 1,\pm 2,\ldots$ is the azimuthal quantum number, and
$j=0,1,2,\ldots$ is the axial quantum number.

The eigenproblem (35) can be solved by employing the optimized
perturbation theory [33--35]. For this purpose, we take as an initial
approximation the Hamiltonian
\be
\label{36}
\hat H_0 = -\; \frac{1}{2}\; {\bf\nabla}^2 +
\frac{1}{2}\left ( u^2 r^2 + v^2 z^2\right ) \; ,
\ee
containing two trial parameters, $u$ and $v$, playing the role of
effective dimensionless frequencies. This Hamiltonian possesses the
eigenfunctions
$$
\psi_{nmj}^0(r,\vp,z) =\left [ \frac{2n!\; u^{|m|+1}}{(n+|m|)!}
\right ]^{1/2} \; r^{|m|} \; \exp\left ( -\; \frac{u}{2}\; r^2\right )
L_n^{|m|}\left ( ur^2\right )\times
$$
$$
\times \; \frac{e^{im\vp}}{\sqrt{2\pi}}\;
\left (\frac{v}{\pi}\right )^{1/4} \frac{1}{\sqrt{2^j\; j!}} \;
\exp\left (-\;\frac{v}{2}\; z^2\right )\; H_j(\sqrt{v}\; z) \; ,
$$
where $L_n^m(\cdot)$ is a Laguerre polynomial and $H_j(\cdot)$ is
a Hermit polynomial. Then, invoking the Rayleigh-Schr\"odinger
perturbation theory, we may find the higher approximations. Thus, for
the spectrum, we get the sequence $\{ E_k(g,u,v)\}$ of approximations,
with $k=0,1,2,\ldots$ being an approximation order. For instance, in
the first order,
\be
\label{37}
E_1(g,u,v) = \frac{p}{2}\left ( u + \frac{1}{u}\right ) +
\frac{q}{4}\left ( v + \frac{\nu^2}{v}\right ) +
u\sqrt{v}\; gI_{nmj} \; ,
\ee
where the dependence of the energy on the quantum numbers is not
explicitly marked in order to avoid cumbersome notation; the quantum
numbers enter through the combinations
$$
p\equiv 2n + |m| + 1  \; , \qquad q\equiv 2j + 1
$$
and through the integral
$$
I_{nmj} \equiv \frac{1}{u\sqrt{v}}\; \int | \psi_{nmj}(r,\vp,z)|^4\;
rdr\; d\vp\; dz \; .
$$
Then the trial parameters $u$ and $v$ are to be transformed to 
{\it control functions} $u_k(g)$ and $v_k(g)$, which play here the 
role of control frequencies, such that the sequence 
$\{ E_k(g,u_k(g),v_k(g))\}$ be convergent [33]. This transformation is
accomplished by means of an optimization condition. Limiting ourselves
by the first-order approximation, we may write the optimization 
condition as
\be
\label{38}
\left ( \dlt u \; \frac{\prt}{\prt u} + \dlt v \;
\frac{\prt}{\prt v} \right )\; E_1(g,u,v) = 0 \; .
\ee
In view of expression (37), this gives the equations
\be
\label{39}
p\left ( 1 -\; \frac{1}{u^2}\right ) + \frac{G}{p\nu}\;
\sqrt{\frac{v}{q}} = 0 \; , \qquad
q\left ( 1 -\; \frac{\nu^2}{v^2}\right ) + \frac{uG}{p\nu\sqrt{vq}} = 0
\ee
for the control frequencies $u=u(g)$ and $v=v(g)$, where the notation
\be
\label{40}
G\equiv 2p\sqrt{q}\; I_{nmj}\; g\nu
\ee
is used. Substituting these control functions into Eq. (37), we obtain
the optimized approximant
\be
\label{41}
E(g) \equiv E_1 (g,u(g),v(g))
\ee
for the energy spectrum. Recall that the energy (41) as well as the
control frequencies $u=u_{mnj}$ and $v=v_{nmj}$ depend on quantum 
numbers.

The spectrum (41), for arbitrary values of $g$, is defined by the system
of equations (37) and (39). In the limits of weak coupling and strong 
coupling, we can derive explicit asymptotic expansions. Thus, for weak
coupling, when $g\ra 0$ and $G\ra 0$, the control frequencies are
$$
u\simeq 1 -\; \frac{G}{2p^2(q\nu)^{1/2}} \; , \qquad
v\simeq \nu - \; \frac{G\nu}{2p(q\nu)^{3/2}} \; .
$$
And the energy spectrum is
\be
\label{42}
E \simeq a_0 + a_1 G \qquad (G\ra 0) \; ,
\ee
where
$$
a_0 = p +\frac{q\nu}{2} \; , \qquad a_1 =\frac{1}{2p(q\nu)^{1/2}} \; .
$$
In the limit of strong coupling, the control frequencies are
$$
u\simeq \frac{p}{G^{2/5}} \; , \qquad
v \simeq \frac{q\nu^2}{G^{2/5}} \; .
$$
And for the energies, we find
\be
\label{43}
E \simeq b_0 G^{2/5} + b_1 G^{-2/5} \qquad (G\ra\infty) \; ,
\ee
where
$$
b_0 = \frac{5}{4} \; , \qquad b_1 = \frac{2p^2+(q\nu)^2}{4}\; .
$$
The difference between two selected energy levels defines the 
transition frequency (18).

If we consider transitions between the ground-state level $(n=m=j=0)$
and an excited level, then calculating the internal transition 
amplitudes (22), we meet the integral
\be
\label{44}
J_{nmj} \equiv \frac{1}{u_0\sqrt{v_0}} \; \int |\psi_0(r,\vp,z)|^2
|\psi_{nmj}(r,\vp,z)|^2 \; rdr \; d\vp \; dz \; ,
\ee
in which $u_0\equiv u_{000}(g)$ and $v_0\equiv v_{000}(g)$. For the 
transition amplitudes (22), corresponding to transitions between the
ground state and an excited mode, labelled by the quantum numbers 
$n,\; m$, and $j$, we obtain
$$
\al_{0,nmj} = gu_0\sqrt{v_0}\left ( I_0 - 2J_{nmj}\right ) \; ,
$$
\be
\label{45}
\al_{nmj,0} = gu_0\sqrt{v_0} \left ( \frac{u_{nmj}}{u_0}\;
\sqrt{\frac{v_{nmj}}{v_0}}\; I_{nmj} - 2J_{nmj}\right ) \; ,
\ee
where $I_0\equiv I_{000}$. The first two inequalities in Eq. (30) for
the internal transition amplitudes now read
\be
\label{46}
\left | \frac{\al_{0,nmj}}{\om_{nmj,0}}\right | \ll 1 \; , \qquad
\left | \frac{\al_{nmj,0}}{\om_{0,nmj}}\right | \ll 1 \; .
\ee
Introduce the quantities
\be
\label{47}
\al_{nmj} \equiv \frac{1}{2}(\al_{0,mnj}+\al_{mnj,0}) \; , \qquad
\dlt_{nmj} \equiv \frac{1}{2}(\al_{0,mnj} - \al_{mnj,0}) \; .
\ee
Here $\al_{nmj}$ is an average transition amplitude, and $\dlt_{nmj}$ 
plays the role of an internal detuning induced by atomic interactions.
By direct numerical calculations of Eqs. (45) and (47), we have checked
that $|\dlt_{nmj}/\al_{nmj}|\leq 0.1$. In appendix, we demonstrate 
calculations for several first excited modes.

The analysis of inequalities (46) shows that their validity imposes a
constraint on the renormalized coupling (40), requiring that the latter
be outside the region of convergence of the strong-coupling expansion
(43). Explicitly, this yields the condition
\be
\label{48}
| g\nu | \leq \frac{[2p^2 +(q\nu)^2]^{5/4}}{14p\sqrt{q}\; I_{nmj}}\; .
\ee
The constraint (48) on the coupling parameter (33) implies that the
number of particles in the trap has to be less than the limiting 
number
\be
\label{49}
N_0 = \frac{[2p^2+(q\nu)^2]^{5/4}}{56\pi\nu p\sqrt{q}\; I_{nmj}}
\left | \frac{l_r}{a_s}\right | \; .
\ee
The value of $N_0$ depends on the quantum numbers, thus, being different
for different modes. For the ground state, Eq. (48) becomes
$$
| g\nu| \leq \left ( 2 + \nu^2\right )^{5/4} \; ,
$$ 
and, respectively, for the limiting number (49), we get
\be
\label{50}
N_0 = \sqrt{\frac{\pi}{2}}\; \frac{(2+\nu^2)^{5/4}}{14\nu}
\left | \frac{l_r}{a_s}\right | \; .
\ee
The limiting number increases for the higher excited modes. Thus, 
it grows as
$$
N_0 \sim (2n+|m|+1)^{3/2} \; , \qquad N_0 \sim (2j+1)^2
$$
according to whether the radial or axial modes are excited. For highly
excited modes, one could invoke an optimized expansion in powers of
$\hbar$ [36]. The number $N_0$ is also very sensitive to the trap shape,
depending on the aspect ratio (31). For the disk-shape $(\nu\gg 1)$ and
cigar-shape $(\nu\ll 1)$ traps, $N_0$ is larger than for a spherical 
$(\nu=1)$ trap.

It is interesting that the limiting number of particles for an excited
coherent mode is close to the critical number of atoms for the stability 
of a trapped Bose gas with attractive interactions. Hence, such gases can 
also support coherent modes. Examples of atoms with a negative scattering 
length are $^7$Li (see review [37]) and $^{85}$Rb (see Ref. [38]). The recent
developments in utilizing Feshbach resonances have shown the ability to 
tune the scattering length to almost any desired value, including the 
ability of changing its sign [3], which makes it possible to add to these 
two species many other atoms. The critical number of trapped Bose-condensed 
atoms with attractive interactions has been estimated earlier [4,39--41]. 
Such trapped atoms form a metastable state, which decays because of quantum 
tunneling, which however, is very slow [42--44], if the number of atoms 
is less than the critical number. There was a proposal [45] for stabilizing 
Bose condensate with attractive interactions by driving a quadrupole 
collective excitation. Our consideration above suggests that it could, 
probably, be also possible to stabilize such condensates by transferring 
them in an excited coherent mode with the help of the coherent resonance.

To estimate the limiting number of particles in a coherent mode, 
consider a cigar-shape trap, as was used [46] for condensing $^{23}$Na.
With the radial frequency $\om_r=1527$ Hz and axial frequency $\om_z=11$ 
Hz, the aspect ratio is $\nu=0.007$. The scattering length of $^{23}$Na
is $a_s=4.498\times 10^{-7}$ cm. Since the oscillator length in this
case is $l_r=1.345\times 10^{-4}$ cm, then $l_r/a_s=300$. The coupling
parameter has to be $g\leq 340$. Therefore, the limiting number 
$N_0\sim 10^4$. If one would take a very long cigar-shape trap with
the aspect ratio $\nu=0.001$, then $g\leq 2400$ and the limiting number
could be as large as $N_0\sim 10^5$.

\section{Phase Portrait}

To study the dynamical behaviour of the system under coherent resonance,
it is convenient to use the population difference
\be
\label{51}
s\equiv |c_2|^2 - |c_1|^2 \; ,
\ee
which varies in the interval $-1\leq s\leq 1$. Then, in view of the
normalization condition (28), we have
$$
|c_1|^2  = \frac{1-s}{2} \; , \qquad |c_2|^2 = \frac{1+s}{2} \; .
$$
Taking this into account, we may write
\be
\label{52}
c_1 =\sqrt{\frac{1-s}{2}}\; \exp\left\{ i\left (\pi_1 +
\frac{\Dlt\om}{2}\; t\right ) \right \} \; , \qquad
c_2 =\sqrt{\frac{1+s}{2}}\; \exp\left\{ i\left (\pi_2 -\;
\frac{\Dlt\om}{2}\; t\right ) \right \} \; ,
\ee
where $\pi_1=\pi_1(t)$ and $\pi_2=\pi_2(t)$ are real-valued phases.

To simplify the notation, we introduce the average interaction amplitude
\be
\label{53}
\al \equiv \frac{1}{2}\left ( \al_{12} +\al_{21}\right ) \; ,
\ee
which is real. Noticing that the transition amplitude (23), due to the
resonant field, is, in general, complex-valued, we can present it as
\be
\label{54}
\bt_{12} = \bt e^{i\gm} \; , \qquad \bt\equiv |\bt_{12}| \; .
\ee
Also, we define the effective detuning
\be
\label{55}
\dlt \equiv \Dlt\om + \frac{1}{2}
\left ( \al_{12} - \al_{21}\right ) \; .
\ee
And, finally, we shall need the phase difference
\be
\label{56}
x \equiv \pi_2 -\pi_1 +\gm \; ,
\ee
whose initial value $x_0$ can be varied by changing the spatial 
dependence of the resonant field. With this notation, Eqs. (29) 
transform to the system of equations for the population difference
\be
\label{57}
\frac{ds}{dt} = -\bt\; \sqrt{1-s^2}\; \sin x
\ee
and for the phase difference
\be
\label{58}
\frac{dx}{dt} = \al s + \frac{\bt s}{\sqrt{1-s^2}}\;\cos x + \dlt \; .
\ee
These are the equations that can be written in the Hamiltonian form
$$
\frac{ds}{dt} = -\; \frac{\prt H}{\prt x} \; , \qquad
\frac{dx}{dt} = \frac{\prt H}{\prt s} \; ,
$$
with the Hamiltonian
\be
\label{59}
H(s,x) = \frac{\al}{2}\; s^2 -\bt\;\sqrt{1-s^2}\; \cos x +
\dlt s \; .
\ee

Let us consider the stationary solutions to Eqs. (57) and (58) in the
rectangle of the variables $-1\leq s\leq 1$ and $0\leq x\leq 2\pi$. 
These solutions are given by the equations
\be
\label{60}
s^4 + 2\ep s^3 - \left ( 1-b^2-\ep^2 \right ) s^2 - 2\ep s -
\ep^2 = 0 \; , \qquad \sin x = 0 \; ,
\ee
where the notation
\be
\label{61}
b \equiv \frac{\bt}{\al} \; , \qquad \ep \equiv \frac{\dlt}{\al}
\ee
is used. In what follows, for simplifying formulas, the dimensionless
detuning $\ep$ will be treated as a small parameter, $|\ep|\ll 1$.
As we have checked by direct numerical calculations, the quantity 
$(\al_{12}-\al_{21})/\al$ is small, and in addition, it can always be
compensated by $\Dlt\om$, so that $\dlt/\al$ be as small as required.

The phase-portrait picture depends on the value of the parameter $b$ 
characterizing the amplitude of the resonant field. This parameter, 
according to notation (61), can be either positive or negative, depending 
on the sign of the interaction amplitude $\al$. The latter can be
negative in the case of attractive interactions. Also, it can become 
negative, as follows from definition (22), even for repulsive 
interactions, when the resonantly connected levels are such that
$$
2\left (|\vp_m|^2,\; |\vp_n|^2\right ) < 
\left ( |\vp_m|^2,\; |\vp_m|^2\right ) \; .
$$
We calculated numerically the scalar products entering the above
inequality and found that, when exciting the lower modes, the amplitude
$\al$ does not change its sign. For this to occur, the resonantly
connected modes must be energetically strongly separated.

When $b^2\geq 1$, there are three  stationary solutions to Eqs. (57) 
and (58),
$$
s_1^* =\frac{\ep}{b}= s_3^* \; , \qquad x_1^*=0\; ,
\qquad x_3^*=2\pi\; ,
$$
\be
\label{62}
s_2^* = -\; \frac{\ep}{b} \; , \qquad x_2^* = \pi \; .
\ee
For $0\leq b<1$, in addition to the stationary points (62), there 
appear two other fixed points,
$$
s_4^*=\sqrt{1-b^2} + \frac{b^2\ep}{1-b^2}\; , \qquad x_4^*=\pi \; ,
$$
\be
\label{63}
s_5^*= -\sqrt{1-b^2} + \frac{b^2\ep}{1-b^2}\; , \qquad x_5^* =\pi\; .
\ee
The points (62) and (63) are the solutions to Eq. (60), which are 
simplified taking account of small detuning $|\ep|\ll 1$. When 
$-1<b\leq 0$, the phase portrait contains seven fixed points, those 
given by Eq. (62) plus
$$
s_4^*=s_6^*\; , \qquad x_4^*= 0 \; , \qquad x_6^* =2\pi\; ,
$$
\be
\label{64}
s_5^*=s_7^*\; , \qquad x_5^*= 0 \; , \qquad x_7^* =2\pi\; ,
\ee
where $s_4^*$ and $s_5^*$ are the same as in Eq. (63). A qualitative
change of a phase portrait is what in dynamical theory called a dynamical
phase transition. In our case, as follows from Eqs. (62) to (64), there
are such dynamical phase transitions at the values of $b=0$ and $b^2=1$.
One may notice that the phase portraits for $b>0$ and $b<0$ are similar 
to each other, just being shifted with respect to $x$ by $\pi$.

The Jacobian matrix associated with Eqs. (57) and (58) is composed of
the elements
$$
J_{11}=\frac{bs}{\sqrt{1-s^2}}\; \sin x \; , \qquad 
J_{12}=-b\;\sqrt {1-s^2}\; \cos x \; ,
$$
$$
J_{21}=1+ \frac{b\cos x}{(1-s^2)^{3/2}}\; , \qquad 
J_{22}=-\; \frac{bs}{\sqrt {1-s^2}}\; \sin x \; .
$$
The eigenvalues of this matrix are given by the equation
\be
\label{65}
J^2 = \frac{b^2}{1-s^2} \left ( s^2\sin^2 x - \cos^2 x\right ) -
b\; \sqrt{1-s^2}\; \cos x \; .
\ee
These eigenvalues are to be evaluated at the fixed points listed above. 
Then for $b^2\geq 1$, we have
\be
\label{66}
J_1^\pm = \pm i\; \sqrt{b(b+1)} = J_3^\pm \; , \qquad
J_2^\pm = \pm i\; \sqrt{b(b-1)} \; ,
\ee
which shows that all fixed points (62) are the centers. When $0\leq b<1$,
then
\be
\label{67}
J_1^\pm = \pm i\; \sqrt{b(1+b)} = J_3^\pm \; , \qquad
J_2^\pm = \pm \sqrt{b(1-b)}  \; , \qquad
J_4^\pm = \pm i\; \sqrt{1-b^2} = J_5^\pm \; , 
\ee
from where if follows that the fixed point $(s_2^*,x_2^*)$ is a saddle, 
while all other points in Eqs. (62) and (63) are the centers. For
$-1<b\leq 0$, the fixed points (64) are similar to those (63), being
shifted by $\pi$ with respect to $x$. All points remain the centers,
except $(s_2^*,x_2^*)$ which is a saddle.

The trajectory, passing through a saddle point, is the saddle separatrix,
which separates the phase plane onto qualitatively different regions of
motion. In the considered case, the separatrix is given by the equality
\be
\label{68}
H(s,x) = H(s_2^*,x_2^*) \; ,
\ee
which, in the approximation linear in $\ep$, results in the separatrix 
equation
\be
\label{69}
\frac{s^2}{2} - b\; \sqrt{1-s^2}\; \cos x + \ep s =  b \; .
\ee
This equation defines two separatrices the upper and lower ones. If the 
initial point $(s_0,x_0)$ of a trajectory lies below the lower separatrix, 
then the variation of $s(t)$ is always limited from above. In the same
way, if the starting point of a trajectory is above the upper separatrix, 
then the change of $s(t)$ is always limited from below by the separatrix.
Since varying the parameter $b$ moves the separatrices, which is clear 
from the separatrix equation (69), the whole phase picture essentially 
changes. The transformation of the phase portrait with varying the
parameter $b$, related to the increase of the transition amplitude due
to the resonant field, is illustrated in Fig. 1, where only the most 
characteristic pictures are presented. The detuning $\ep$ in these
figures is set zero. A finite detuning $\ep\neq 0$ makes the pictures
slightly asymetric with respect to the line $s=0$, but the overall phase
portraits are similar, because of which we do not present them here.

The structure of the phase portraits shows that, at small $b$, there are
a number of trajectories always lying in the upper part of the plane if
their initial points were above the upper separatrix. In particular, if
the upper coherent mode was initially highly populated, and the applied 
resonant field is weak, the mode will retain its high population. In the
extreme case, if at the initial time all atoms are in the upper coherent
mode, and there is no external field, so that $s_0=1$ and $b=0$, then
$s(t)=1$ meaning that atoms remain in the upper mode. Such a setup can 
be achieved in two ways. For instance, by transferring atoms to the upper
level through an intermediate third higher mode. Or, even simpler, just
by applying, first, a resonant field sufficient for transferring atoms
from the ground state to the chosen coherent mode and then switching 
off this field. Thus atoms remain locked in this upper mode. This {\it 
mode-locking effect} was described in Ref. [4]. Mathematically, it 
is analogous to the self-trapping of atoms in one of the wells of a 
stationary double-well potential [30--32,47]. However, the physics of 
the coherent resonance occurring under the action of a resonant external
field is absolutely different. Moreover, the mode-locking effect is not
the most interesting thing happening in the process of the resonant  
excitation of coherent modes.

Varying the amplitude of the resonant field, that is, the parameter 
$b$, in the dynamic behaviour of fractional populations there appear
{\it dynamical critical phenomena} [19,20], arising on a critical line 
of the parametric manifold $\{ b,\ep\}$. The qualitative behaviour of 
fractional populations drastically changes when the critical line is 
crossed. As was shown [19,20], for a stationary system, obtained by 
time-averaging the considered dynamical system, there appear critical 
phenomena, typical of phase transitions and occurring on the same critical 
line, where the dynamical critical phenomena are observed. Thus, we may 
define a pumping capacity, playing the role of an effective specific heat,
and an effective susceptibility, which diverge on the critical line [19,20].
The origin of these dynamical critical phenomena has not been completely 
understood. But now, invoking the analysis of this section, we clearly see 
what happens.

Suppose, we consider a trajectory starting at the point $\{ s_0,x_0\}$. 
Strengthening the applied resonant field means increasing the parameter $b$. 
But changing $b$ implies moving the separatrix given by Eq. (69). When the 
separatrix crosses the initial point $\{ s_0,x_0\}$, the trajectory passes 
from one region of the phase plane to a qualitatively different region, 
because of which the dynamical behaviour changes qualitatively. Since, in 
general, we have two independent parameters, the amplitude $b$ and detuning 
$\ep$, the {\it effect of separatrix crossing} can be due to the variation 
of any of these parameters. A tiny change of one of these can shift the 
trajectory to a different part of the phase plane. The condition for the 
separatrix crossing defines, on the parametric manifold $\{ b,\ep\}$, the 
{\it critical line} 
\be
\label{70}
\frac{s_0^2}{2} -b_c\;\sqrt{1-s_0^2}\;\cos x_0 +\ep_c s_0 = |b_c|\;.
\ee
Actually, for each set of initial conditions $\{ s_0,x_0\}$ there exists 
its own critical line defined by Eq. (70). For the variety of initial
conditions, there is an infinite bunch of critical lines.

For example, let us consider the initial condition $s_0=-1,\; x_0=0$,
when all atoms at the initial time are in the ground mode. Then the
critical line (70) reduces to
$$
|b_c| +\ep_c = \frac{1}{2} \; .
$$
In the case of zero detuning, this defines two points $b_c=\pm 0.5$. Assume 
that our concern is the behaviour of atoms with repulsive interactions, such 
that $b>0$. And let us set $\ep=0$. The population difference $x(t)$ exhibits 
dramatic changes when $b$ crosses the critical point $b_c=0.5$. This is 
illustrated in Fig. 2, where the change of $b$ around the critical point 
$b_c$ occurs in the fifth decimal digit, so that the variation of $b$ from 
Fig. 2(a) to Fig. 2(b) makes $\Dlt b=0.00001$. Slightly below the critical 
point $b_c$, the population difference $s(t)$ oscillates between $-1$ and 
$0$, while the phase difference $x(t)$ monotonely diminishes. Just above the 
critical point, the oscillation period of $s(t)$ doubles, as well as the 
amplitude of $s(t)$, as now $s(t)$ varies between $-1$ and $+1$. The form of 
$s(t)$ also changes acquiring a specific cusp in the middle of the period. 
The phase difference $x(t)$, instead of being monotonely diminishing, becomes 
oscillating.

If we are interested in the behaviour of atoms with $b<0$, say with 
attractive interactions, then similar dramatic changes happen at the
critical point $b_c=-0.5$. This is shown in Fig. 3. Again, the change
of $b$ around the critical point $b_c$ occurs in the fifth decimal 
digit, as for Fig. 2. After crossing the critical point $b_c$, the 
period and amplitude of $s(t)$ double, and the middle-period cusps 
appear. The shapes of both $s(t)$ and $x(t)$ essentially change. Note 
that although these dynamical critical phenomena for $b>0$ and $b<0$ 
are similar, there is a difference in the behaviour of $x(t)$. For 
$b<0$, the phase $x(t)$ is an oscillating function below as well as 
above the critical point. While for $b>0$, the phase difference $x(t)$ 
below the critical point is a monotonely diminishing function.

In Figs. 2 and 3, time is dimensionless (measured in units of $\al^{-1}$). 
The return to dimensional time is accomplished by the substitution 
$t\ra t/\al$. A small detuning $\ep\neq 0$ slightly curves the lines but 
does not make qualitative change.

\section{Power Broadening}

The feasibility of coherent resonance imposes a limitation on the 
value of the internal transition amplitude, that is, as discussed 
in Sec. IV, on the admissible number of particles. In addition to 
the amplitude restriction, there is one more limitation having to 
do with temporal power broadening. Even if the alternating external 
field is well tuned to the resonance with a chosen transition frequency, 
nevertheless, there exist nonresonant transitions to other levels. 
Though the probability of nonresonant transitions, at each given 
instant of time, is very small, their influence on the process of 
resonant excitation can increase with time, becoming essential for 
long time intervals. To estimate this influence, we return to the 
evolution equations (29).

It is convenient to introduce the quantity
\be
\label{71}
 h\equiv 2c_1^* c_2 e^{i(\Dlt\om t +\gm)} = \sqrt{1-s^2}\; e^{ix}\; ,
\ee
playing the role of a ladder variable in optics [26,28]. Here $s=s(t)$
and $x=x(t)$ are the same as in Eqs. (51) and (56). Define the collective
frequency
\be
\label{72}
\Om \equiv \al s + \dlt \; .
\ee
Here and in what follows, we use the notation for $\al,\; \bt$, and
$\dlt$ as given in Eqs. (53), (54), and (55), respectively. Taking
account of nonresonant transitions can be done by including in the
evolution equations an inhomogeneous broadening modelled by a random 
variable [28]. Then, Eqs. (29) can be rewritten in the form of the
stochastic equations
\be
\label{73}
\frac{dh}{dt} = i\left ( \Om +\xi\right ) h + i\bt s \; ,
\ee
\be
\label{74}
\frac{ds}{dt} = \frac{i}{2}\; \bt \left ( h - h^*\right ) \; ,
\ee
where $\xi=\xi(t)$ is a stochastic variable modelling nonresonant 
transitions.

In analysing Eqs. (73) and (74), we assume, for simplicity, that the
transition amplitude due to the external field is weaker than that
caused by internal atomic interactions,
\be
\label{75}
\left | \frac{\bt}{\al}\right | \ll 1 \; .
\ee
The stochastic variable can be presented as Gaussian white noise [48],
with the stochastic averages
\be
\label{76}
\ll \xi(t)\gg \; = 0 \; , \qquad
\ll \xi(t) \; \xi(t') \gg \; = 2\gm_3\; \dlt(t-t') \; .
\ee
The noise width can be approximated as
\be
\label{77}
\gm_3 \approx \frac{\al^2 + \bt^2}{\om_{21}} \; ,
\ee
so that, according to inequalities (30), $\gm_3\ll\om_{21}$.

Equations (73) and (74) can be solved by employing the scale separation 
approach [49--51], which is a generalization of the averaging technique
[27,52] to stochastic differential equations. Because of inequality 
(75), the function $h=h(t)$ can be treated as fast, compared to the slow
function $s=s(t)$. The latter is a temporal quasi-invariant with respect
to $h$. In this case, the solution to Eq. (73) reads
\be
\label{78}
h = h_0 e^{i\Om t} \exp\left\{ i \int_0^t \xi(t')\; dt'\right \} +
i\bt s \int_0^t e^{i\Om(t-t')} \exp\left\{ i\int_{t'}^t \xi(t'')\;
dt'' \right\} \; dt'\; ,
\ee
where $h_0\equiv h(0)$. Averaging Eq. (78) over the stochastic variable,
with the use of Eqs. (76), we have
\be
\label{79}
\ll h\gg \; = \left ( h_0 + \frac{\bt s}{\Om +i\gm_3}\right )
e^{(i\Om-\gm_3)t} - \; \frac{\bt s}{\Om+i\gm_3} \; .
\ee
Substituting the solution (78) for the fast function into Eq. (74) for 
the slow function and averaging the right-hand side of Eq. (74) over
the stochastic variable and over time, we obtain the equation
\be
\label{80}
\frac{ds}{dt} = -\; \frac{\bt^2\gm_3 s}{(\al s+\dlt)^2 + \gm_3^2}
\ee
for the guiding center of the slow function. An approximate solution
to this equation can be written as
\be
\label{81}
s \approx s_0 \exp\left ( - \; \frac{\bt^2\gm_3 t}{\Om^2 +\gm_3^2}
\right ) \; .
\ee
This shows that the population difference (81) attenuates, with the
characteristic time
\be
\label{82}
t_c \equiv \frac{(\al s_0 +\dlt)^2 +\gm_3^2}{\bt^2 \gm_3} \; .
\ee
From here it follows that the process of resonant pumping is not 
essentially influenced by power broadening only during times smaller 
than $t_c$. After the time (82), the two-level resonant picture will
not be adequate, since the neighboring nonresonant levels will be
essentially involved in the process. The characteristic time (82), in 
view of Eq. (77), is
\be
\label{83}
t_c \cong \frac{(\al s_0+\dlt)^2\om_{21}^2 +(\al^2+\bt^2)^2}
{(\al^2+\bt^2)\bt^2\om_{21}} \; .
\ee
Taking account of the inequalities $|\dlt/\al|\ll 1$ and 
$|\bt/\al|\ll 1$, and setting $|s_0|\approx 1$, simplifies the time 
(83) to
\be
\label{84}
t_c \simeq \frac{\al^2 +\om_{21}^2}{\bt^2\om_{21}} \; .
\ee
Finally, under conditions (46), Eq. (84) reduces to
\be
\label{85}
t_c \simeq \frac{\om_{21}}{\bt^2} \; .
\ee
To estimate $t_c$ we may take $\bt\sim 0.1\al$ and $\al\sim 0.1\om_{21}$,
with a transition frequency, typical of magnetic traps [3], of order
$\om_{21}\sim 10^2 - 10^3$ Hz. This gives $t_c\sim 10-100$ s, which is
a rather long time, comparable or longer than the lifetime of atoms 
inside a trap [2,3].

In this way, the temporal limitation resulting from the power broadening
does not impose too severe restrictions on the procedure of the resonant
excitation of coherent modes. Moreover, as is discussed in Sec. V, we do 
not need to pump too long but we may stop pumping as soon as the desired
mode is populated, which happens during the time of order $2\pi/\Om$.
The latter is of order $0.1$ s. Hence, the time necessary for populating
a chosen coherent mode is essentially less than the critical time $t_c$, 
after which power broadening would spoil the resonant picture.

\section{Interference Patterns}

In the process of the resonant excitation of coherent modes, only two 
modes are involved in the dynamical picture, while the population of
all other nonresonant modes remains negligibly small. This allows us to 
present the system coherent function as the sum
\be
\label{86}
\vp(\br,t) = \vp_1(\br,t) + \vp_2(\br,t)
\ee
of the terms
\be
\label{87}
\vp_i(\br,t) =c_i(t)\vp_i(\br) \exp \left ( -\; \frac{i}{\hbar}\;
E_i t\right ) 
\ee
corresponding to the related modes $i=1,2$. From here it follows that 
there should exist a spatial interference between the modes, similarly
to the interference of different atomic components in a binary mixture
of two Bose condensates [53]. In our case, the interference becomes 
possible because different coherent modes possess qualitatively
different spatial shapes. Thus, we may define [3] the {\it interference
pattern}
\be
\label{88}
\rho_{int}(\br,t) \equiv \rho(\br,t) - \rho_1(\br,t) -\rho_2(\br,t)\; ,
\ee
in which
\be
\label{89}
\rho(\br,t) \equiv |\vp(\br,t)|^2 \; , \qquad
\rho_i(\br,t) \equiv |\vp_i(\br,t)|^2 \; .
\ee
With the use of the dimensionless real function
\be
\label{90}
\psi_i(r,z) \equiv l_r^{3/2} \vp_i(\br) e^{-im_i\vp} \; ,
\ee
in which $l_r$ is the oscillator length and $m_i$ is a winding number, 
the pattern (88) reads
\be
\label{91}
\rho_{int}(\br,t) = \frac{\psi_1\psi_2}{l_r^3} \; \sqrt{1-s^2}\;
\cos\Phi \; ,
\ee
where $s=s(t)$ is defined by Eq. (51) and the notation
\be
\label{92}
\Phi \equiv (m_2-m_1)\vp + x(t) -\gm -(\om_{21}+\Dlt\om) t
\ee
is introduced. This interference pattern can be experimentally observed 
either inside the trap by means of light scattering or by freeing atoms
from the trap and observing their free evolution, as is discussed by
Sinatra and Castin [53].

The existence of two modes in a trap, with different spatial shapes 
of the related wave functions, leads to the appearance of atomic 
current inside the trap. This effect is analogous to the Josephson 
effect, which is usually considered for two atomic clouds sitting in 
a double-well potential, so that the clouds are separated by a potential
barrier [29--32]. However, as is suggested by Leggett [54], Josephson 
oscillations can exist between two interpenetrating populations, not
separated by any barrier. In this case, one calls it the internal 
Josephson effect [55]. Such a tunneling, involving no potential barriers,
is also called quantum dynamical tunneling [56,57], which is actually
just a current between two modes representing bound states.

The {\it interference current} is
\be
\label{93}
{\bf j}_{int}(\br,t)  \equiv {\bf j}(\br,t)  - {\bf j}_1(\br,t)  -
{\bf j}_2(\br,t)  \; ,
\ee
where
\be
\label{94}
{\bf j}(\br,t) = \frac{\hbar}{m_0} \; {\rm Im}\; \vp^*(\br,t)
{\bf\nabla}^2\vp(\br,t) \; , \qquad
{\bf j}_i(\br,t)  =\frac{\hbar}{m_0}\; {\rm Im}\; \vp_i^*(\br,t)
{\bf\nabla}\vp_i(\br,t) \; ,
\ee
with $i=1,2$. Employing the notation in Eqs. (90) and (92), we find
\be
\label{95}
{\bf j}_{int}(\br,t) = \frac{\hbar\psi_1\psi_2}{2m_0l_r^4}\;
\sqrt{1-s^2} \left [ \left ( {\bf e}_r\; \frac{\prt}{\prt r} +
{\bf e}_z\; \frac{\prt}{\prt z}\right ) 
\ln\left (\frac{\psi_2}{\psi_1}\right ) \sin\Phi + {\bf e}_\vp \;
\frac{m_1+m_2}{r}\; \cos\Phi \right ] \; .
\ee

Both the interference patterns (91) and interference current (95)
experience temporal oscillations on two time scales. One corresponding
to the transition frequency $\om_{21}$, which is typical of the 
Josephson effect. And another scale is related to the temporal variation
of $s(t)$ and $x(t)$, which is conneted with the transition amplitudes
$\al$ and $\bt$. Since $\al,\bt\ll\om_{21}$, the Josephson oscillations 
are fast, as compared to the slow change of $s(t)$ and $x(t)$. The fast
Josephson oscillations, being modulated by the slow variation of $s(t)$
and $x(t)$, yield the collapse-revival picture typical of some two-level 
systems [26]. Here the modulation is mainly due to the temporal 
variation of $s(t)$. If one stops applying the resonant field, so that
$\bt=0$, then $s(t)=const$, and the slow modulation disappears.

\section{Spin Squeezing}

The considered atomic system is essentially nonlinear. Therefore, one might 
expect that some kind of squeezing effects could arise. Similarly to squeezed 
states of light [58,59], one can introduce squeezed atomic states [58]. This 
is usually considered for the case of two-level atoms, each of which possesses
two internal states. Finite level systems, as is known, are conveniently 
described by means of spin operators, because of which atomic squeezing is 
commonly called spin squeezing. In order to emphasize that the spin operators,
employed for describing finite-level atoms, are not actually the operators 
representing real spins, but rather are convenient mathematical tools, one 
also uses the terms of dipole squeezing [58,60] or pseudospin squeezing [61]. 
In general, one may define squeezing for other operators from a Lie algebra 
[61,62]. Atomic squeezing is directly related to the radiation field squeezing
[63]. And vice versa, squeezed atomic states can be created by irradiating 
atoms with squeezed light [64--67] or with light combined with an alternating 
magnetic field, as in the process of continuos quantum measurement [68]. 
Generally, atomic squeezing can be achieved for both bosons as well as for 
fermions [69,70]. Squeezing in Bose-Einstein condensates was considered for 
two-component mixtures [71], for atoms with two internal states [72], which 
is equivalent to a two-component mixture, and for atoms in linked mesoscopic 
traps formed by an optical lattice [73], which is equivalent to a 
multicomponent mixture. There is a real potential for several practical 
applications of squeezed atoms, e.g., for atomic spectroscopy and atomic 
clocks [74], for atom interferometers [75], and, probably, for quantum 
computation [76].

To consider spin squeezing in our case, we may notice that the transition 
dipole (71) and population difference (51) can be presented as the statistical
averages 
\be
\label{96}
h \equiv \frac{2}{N}\; < S_->\; , \qquad s\equiv \frac{2}{n}\;
<S_z>
\ee
of the collective spin operators
\be
\label{97}
S_\al \equiv \sum_{i=1}^N S_i^\al \; , \qquad
S_\pm \equiv S_x \pm i S_y \; ,
\ee
where $\al=x,y,z$ and $S_i^\al$ corresponds to a $1/2$-spin operator.
The evolution equations (73) and (74) can be obtained by averaging the
Heisenberg equations of motion for the spin operators (97), with the
effective Hamiltonian
\be
\label{98}
H_{eff} = \frac{\bt}{2}\left ( S_- + S_+ \right ) - \dlt S_z -\;
\frac{\al}{N} \; S_z^2 \; .
\ee

Atomic squeezing is defined in a way similar to the squeezing of light,
being based on the quantum-mechanical uncertainty relations. For any
two operators $A$ and $B$, not necessarily Hermitian, the Heisenberg
uncertainty relation reads
\be
\label{99}
\Dlt^2(A) \; \Dlt^2(B) \geq \frac{1}{4}\;
\left | <[A, \; B]>\right |^2 \; ,
\ee
where the dispersion is
\be
\label{100}
\Dlt^2(A) \equiv <A^+A>\; - \; |<A>|^2 \; .
\ee
One says that $A$ is squeezed with respect to $B$ if
$$
\Dlt^2(A) < \frac{1}{2}\; |<[A,\; B]>| \; .
$$
This suggests to introduce the {\it squeezing factor}
\be
\label{101}
Q_{AB} \equiv \frac{2\Dlt^2(A)}{|<[A,\; B]>|} \; .
\ee
Now, for $A$ and $B$, we may take any spin operators. However, taking 
separately $S_\al$, with $\al=x,y,z$, may involve the so-called trivial
squeezing due to rotation [77]. To avoid this, we consider the operators
$S_z$ and $S_\pm$, for which we have
\be
\label{102}
\Dlt^2(S_z)\; \Dlt^2(S_\pm) \geq \frac{1}{4}\; |<S_\pm>|^2 \; .
\ee
Squeezing of $S_z$, with respect to $S_\pm$, is defined by the squeezing
factor $Q_{S_z S_\pm} \equiv Q_z$, which in view of definition (101), is
\be
\label{103}
Q_z = \frac{2\Dlt^2(S_z)}{|<S_\pm>|} \; .
\ee
The squeezing factor (103), taking into account that 
$|<S_\pm>|^2=<S_x>^2+<S_y>^2$, becomes
\be
\label{104}
Q_z = \frac{2\Dlt^2(S_z)}{\sqrt{<S_x>^2+<S_y>^2}} \; ,
\ee
which has the form used in Refs. [72,74,77]. One can say that $S_z$ is
squeezed with respect to $S_\pm$ if $\Dlt^2(S_z)<\frac{1}{2}|<S_\pm>|$,
that is, $Q_z<1$. Calculating the expressions
$$
\Dlt^2(S_z) = \frac{N}{4}\; \left ( 1-s^2 \right ) \; , \qquad
\Dlt^2(S_\pm) = \frac{N}{4}\; \left ( 1 \mp s^2\right ) \; , \qquad
|<S_\pm>| = \frac{N}{2}\; |h| =\frac{N}{2}\; \sqrt{1-s^2}\; ,
$$
we obtain the squeezing factor
\be
\label{105}
Q_z = \sqrt{1-s^2} \; ,
\ee
where $s=s(t)$ is the population difference satisfying the evolution 
equations discussed above. As is seen, since $0\leq s^2\leq 1$, the
factor (105) is almost always less than one, except for $s=0$. Hence 
$S_z$ is squeezed with respect to $S_\pm$. In other words, the dispersion 
$\Dlt^2(S_z)$ is almost always smaller than the dispersion $\Dlt^2(S_\pm)$. 
This means that the population difference and, respectively, the fractional 
populations can be measured with a better accuracy than the transition dipole 
or the relative phase between population amplitudes. The characteristic 
temporal behaviour of the squeezing factor (105) is shown in Fig. 4, with 
$s(t)$ found numerically from the evolution equations.

It is worth noting that the treatment of atomic squeezing by means of spin 
operators is often used in the quantum optics language [58--61]. This is
equivalent to the formation of squeezed states generated by the Bogolubov 
canonical transformation for quasiparticles in multicomponent (spinor) 
condensate. These two pictures are equivalent mathematically and are caused 
by the same physical reason, by the existence of nonlinear interactions in 
a multicomponent system.

In the theory of nuclear magnetic resonance [78,79], one distinguishes
the state coherence, when $s=\pm 1$, and the transition coherence, when
$|h|=1$. This terminology can also be applied to our case. Since here 
we have $|h|^2=1-s^2$, then the state and transition coherences are 
complimentary to each other. And if $Q_z<1$, this means that the state 
coherence can be better controlled than the transition coherence.

Atomic squeezing is usually neighbours with atomic entanglement [72,80,81]. 
Entanglement is simply the Schr\"odinger's name for superposition in a 
multiparticle system. In the system of trapped atoms or ions with two or
more internal states, multiparticle entanglement can appear [81]. This 
concerns as well Bose-Einstein condensates composed of atoms having 
internal states [72,82]. The Bose-condensed atomic cloud, subject to the
coherent resonance we study here, is also in a multiparticle entangled 
state. Really, the multiparticle density matrix, for the case studied, is
$$
\rho_N(\br_1,\br_2,\ldots,\br_N,\br_1',\ldots,\br_N',t) = |c_1(t)|^2 
\prod_{i=1}^N \vp_1(\br_i)\;\vp_1^*(\br_i') + |c_2(t)|^2 \prod_{i=1}^N 
\vp_2(\br_i)\;\vp_2^*(\br_i') \; , 
$$
where $|c_1|^2+|c_2|^2=1$. This function cannot be written in any way as a 
product of single-particle functions. Hence $\rho_N$ is an entangled state, 
being a kind of the mixed counterpart of the Greenberger-Horne-Zeilinger 
state [83,84].

The principal difference of the two-mode coherent system, we consider, 
from the entanglement of trapped atoms, studied earlier, is that the
nonlinear coherent modes describe {\it not internal} states of separate
atoms, but {\it collective} coherent states of the whole system.
Therefore this type of entanglement can be called {\it coherent 
entanglement}.

\section{Relaxation Process}

In the previous sections, we have considered the situation when 
Bose-condensed atoms are subject to the permanent pumping by an 
alternating resonant field. One might ask the question, what happens 
if this resonant pumping is stopped? How to describe the behaviour of
the system {\it after} the coherent resonance?

If the resonant field ceases acting on the system, say at the moment of
time $t_0$, then, after this time, the two-mode picture becomes invalid 
and one needs to return back to the initial equation (6) for the coherent
field. This equation is to be complemented by the relaxation terms
characterizing atomic collisions and possible existence of noncondensed
atoms [1--3]. Then the coherent-field equation reads
\be
\label{106}
i\hbar \; \frac{\prt\eta}{\prt t} = H[\eta]\; \eta -\;
\frac{i\hbar}{2}\left ( K_2 |\eta|^2 + K_3|\eta|^4 +\Gm M\right )\;
\eta \; ,
\ee
in which $H[\eta]$ is the nonlinear Hamiltonian (7), $K_2$ and $K_3$ are
two-body and three-body recombination loss-rate coefficients [85], $\Gm$
is the loss rate due to the transfer of condensed atoms to noncondensed
atoms whose number is $M$. If a cooling mechanism is supported, so that
there is an opposite process of transferring atoms from a noncondensed
cloud to the condensate, then $\Gm<0$ is a gain rate. The number of
particles in the condensate is $N=||\eta||^2$. The initial condition for
Eq. (106), in view of relation (8), is prescribed by the function
\be
\label{107}
\eta(\br,t_0) =\sqrt{N}\; \vp(\br,t_0) \; ,
\ee
with $\vp(\br,t)$ given by the sum (86) taken at the moment $t_0$, when
the resonant field stopped acting on the system.

In addition to loss rates caused by depolarizing collisions, there can 
exist another internal natural loss rate
\be
\label{108}
\Gm_1 \equiv -\; \frac{2}{\hbar}\; {\rm Im} \left (\vp,\; 
\hat H[\vp] \; \vp\right )
\ee
appearing for atoms with attractive interactions, if the number of atoms 
is larger than the critical number $N_c$. Here $\hat H[\vp]$ is the
nonlinear Hamiltonian (11). Thus, for a spherical trap, the loss rate
(108) is estimated [4] as
$$
\Gm_1 \simeq 2.867\; \om_r\Theta(N-N_c) \left | \frac{a_s}{l_r}\;
N\right |^{2/5} \; ,
$$
where $N\gg N_c$ and $\Theta(\cdot)$ is a unit step function. The loss
rates due to two-body and three-body depolarizing collisions are
\be
\label{109}
\Gm_2 \equiv K_2\left ( |\vp|^2,\; |\vp|^2 \right ) \; , \qquad
\Gm_3 \equiv K_3\left ( |\vp|^3,\; |\vp|^3 \right )  \; .
\ee
From the evolution equation (106), we can easily derive the rate
equation for condensed atoms, which, being complimented by the rate
equation for noncondensed atoms, makes the general set of equations
\be
\label{110}
\frac{dN}{dt} = -\Gm_1 N - \Gm_2 N^2 - \Gm_3 N^3 - \Gm MN \; , \qquad
\frac{dM}{dt} =-\gm_1( M - M_0 ) + \Gm NM \; ,
\ee
defining the relaxation process in the system after the coherent 
resonance. Here $\gm_1$ is a pumping rate for noncondensed atoms and 
$M_0$ is their stationary number.

It is not our aim to give in this paper a detailed analysis of the
relaxation equations (110). This is a separate problem which could 
be investigated in other publications. Here our main goal has been 
to present a thorough description of trapped Bose-Einstein condensate
under the condition of coherent resonance. Therefore, we limit this 
section by a brief sketch of the way that would allow one to describe 
in detail what happens after the coherent resonance. The relaxation
procedure defined by Eqs. (110) can follow quite different patterns
depending on the concrete physical situation and on the values of the
related relaxation rates. In a particular case, when the natural loss
rate (108) is zero and the are no noncondensed atoms, the relaxation 
is completely due to depolarizing collisions. Then the relaxation time 
is the lifetime of atoms in a trap. For different traps this time 
varies between $10$ s and $100$ s. Note that the loss rate (108) is zero 
for atoms with repulsive interactions and also for atoms with attractive
interactions, if the number of atoms is less than critical. In such 
cases, nonlinear coherent modes, created by means of coherent resonance,
can live, after the pumping resonance field is switched off, quite long 
time, of the order of the lifetime of atoms in a trap. This makes it 
feasible to study their behaviour as well as to use them for practical
applications.

\section{Conclusions}

We have demonstrated that nonlinear coherent modes of trapped 
Bose-condensed atoms can be created by means of a resonant alternating 
field. These modes represent collective states corresponding to 
nonground-state Bose-Einstein condensates. Conditions, when such 
a resonant excitation of coherent modes is possible, are investigated. 
One restriction is that the number of atoms in a mode be less than 
a limiting number. The latter depends on the type of atoms and trap 
characteristics, and can be as large as $10^5$. This limiting number 
is close to that required for the stability of atoms with negative 
scattering lengths. Hence, atoms with attractive interactions can also
be employed for creating nonlinear coherent modes. Moreover, since the 
mode limiting number increases for higher modes, it could be possible
to stabilize a larger number of atoms with negative scattering length
by transferring them to such excited coherent modes.

Another restriction on the coherent resonance is that the resonant 
pumping can last not longer than a critical time, before power 
broadening spoils the resonant picture. However, this limitation is 
not dangerous because of two reasons. First, the critical time for 
power broadening is rather long, of order $10-100$ s, which is about 
the lifetime of atoms in a trap. Second, there is no need to pump the 
system for so long times, since the transfer to an excited coherent mode
takes essentially shorter times, around $0.1$ s.

Temporal behaviour of fractional populations, in the process of 
coherent resonance, exhibits dynamic critical phenomena occurring on 
a critical line in the parametric manifold. The related time-averaged 
system displays, on this critical line, critical effects typical of 
phase transitions. The origin of these critical phenomena is the saddle
separatrix crossing by a starting point of a trajectory.

Interference patterns and interference current can be observed. These 
are related to the internal Josephson effect and to dynamic barrierless 
tunneling.

Atomic squeezing is realized demonstrating that the state coherence 
is better defined than transition coherence. Massive multiparticle 
entanglement is produced. A specific feature of this coherent 
entanglement, distinguishing it from earlier studies, is that it
involves not internal states of individual atoms but collective
coherent states of atomic condensate.

After the pumping resonant field is switched off, the behaviour of the
system can be described by the rate equations. This behaviour depends 
on the physical setup defining the related relaxation rates. When the
loss of atoms in a nonlinear coherent mode is caused by depolarizing 
atomic collisions, the lifetime of the mode is about $10-100$ s, which 
is the lifetime of atoms in a trap.

The two-mode picture of a resonant Bose condensate somewhat resembles 
the two-level picture of a resonant atom. However there is a principal
difference between these two cases in the nature of the states involved. 
The notion of a resonant atom involves internal states of a single atom,
while that of a {\it resonant condensate} has to do with collective coherent
states of the whole system. Thus, the resonant condensate is a principally 
new resonant physical system. Its similarity with resonant atoms makes 
it feasible to extend to this type of systems various applications that
are elaborated for resonant atoms. And its difference from the latter
brings hopes of discovering new effects and finding novel applications.

\vskip 5mm

{\bf Acknowledgement}

\vskip 2mm

One of the authors (V. I. Y.) is very grateful for discussions and 
useful comments to M. D. Girardeau, V. K. Melnikov, and E. Zaremba.

The work has been accomplished in the Research Center for Optics and 
Photonics, University of S\~ao Paulo, S\~ao Carlos. Financial support
from the S\~ao Paulo State Research Foundation (Fapesp) is appreciated.

\newpage

{\Large{\bf Appendix}}

\vskip 3mm

To illustrate the calculations of Sec. IV, we present here the 
related explicit expressions for several first nonlinear coherent 
modes, including the ground state mode $(n=m=j=0; \; p=q=1)$, radial 
dipole mode $(n=1,\; m=j=0; \; p=3,\; q=1)$, basic vortex mode 
$(n=0,\; m=1,\; j=0; \; p=2,\; q=1)$, and axial dipole mode
$(n=m=0,\; j=1; \; p=1,\; q=3)$. The corresponding wave functions are
$$
\psi_0(r,\vp,z) =\left ( \frac{u_0^2v_0}{\pi^3}\right )^{1/4}\;
\exp\left \{ -\; \frac{1}{2}\left ( u_0r^2 + v_0 z^2\right )
\right\} \; ,
$$
$$
\psi_{100}(r,\vp,z) =\left ( \frac{u_{100}^2v_{100}}{\pi^3}
\right )^{1/4}\; \left (u_{100}r^2 -1 \right )\;
\exp\left \{ -\; \frac{1}{2}
\left ( u_{100}r^2 + v_{100} z^2\right ) \right\} \; ,
$$
$$
\psi_{010}(r,\vp,z) =u_{010}\; \left ( \frac{v_{010}}{\pi^3}
\right )^{1/4}\; re^{i\vp} \; \exp\left \{ -\; \frac{1}{2}
\left ( u_{010}r^2 + v_{010} z^2\right ) \right\} \; ,
$$
$$
\psi_{001}(r,\vp,z) =\left ( \frac{4u_{001}^2v_{001}^3}{\pi^3}
\right )^{1/4}\; z\; \exp\left \{ -\; \frac{1}{2}
\left ( u_{001}r^2 + v_{001} z^2\right ) \right\} \; ,
$$
where $u_{nmj}$ and $v_{nmj}$ are the control functions defined by 
Eq. (39).

The corresponding integrals $I_{nmj}$ are
$$
I_{000} =(2\pi)^{-3/2} \equiv I_0 = 0.063494\; , \qquad
I_{100}=\frac{1}{2}\; I_0 \; , \qquad I_{010} = \frac{1}{2}\; I_0\; ,
\qquad I_{001} = \frac{3}{4}\; I_0 \; .
$$
And the integrals (44) are
$$
J_{100} = \frac{u_{100}(u_0^2+u_{100}^2)}{\pi^{3/2}(u_0+u_{100})^3} 
\left ( \frac{v_{100}}{v_0+v_{100}}\right )^{1/2} \; , \qquad
J_{010} = \frac{u_{010}^2}{\pi^{3/2}(u_0+u_{010})^2} 
\left ( \frac{v_{010}}{v_0+v_{010}}\right )^{1/2} \; , 
$$
$$
J_{001} = \frac{u_{001}}{\pi^{3/2}(u_0+u_{001})} 
\left ( \frac{v_{001}}{v_0+v_{001}}\right )^{3/2} \; .
$$

In the weak-coupling limits, when $g\sqrt{\nu}\ra 0$, the control 
functions behave as
$$
u_0\simeq 1- I_0 g \sqrt{\nu} \; , \qquad 
v_0\simeq \nu - I_0 g\sqrt{\nu} \; , 
$$
$$
u_{100} \simeq 1- \; \frac{1}{6}\; I_0 g \sqrt{\nu} \; , \qquad 
v_{100} \simeq \nu - \; \frac{1}{2}\; I_0 g\sqrt{\nu} \; , 
$$
$$
u_{010} \simeq 1- \; \frac{1}{4}\; I_0 g \sqrt{\nu} \; , \qquad 
v_{010} \simeq \nu - \; \frac{1}{2}\; I_0 g\sqrt{\nu} \; , 
$$
$$
u_{001} \simeq 1- \; \frac{3}{4}\; I_0 g \sqrt{\nu} \; , \qquad 
v_{001} \simeq \nu - \; \frac{1}{4}\; I_0 g\sqrt{\nu} \; .
$$
And for the energy spectrum (42), we have
$$
E_0 \simeq 1 + \frac{\nu}{2} + I_0 g\sqrt{\nu} \; , \qquad
E_{100} \simeq 3 + \frac{\nu}{2} + \frac{1}{2}\; I_0 g\sqrt{\nu} \; ,
$$
$$
E_{010} \simeq 2 + \frac{\nu}{2} + \frac{1}{2}\; I_0 g\sqrt{\nu} \; ,
\qquad
E_{001} \simeq 1 + \frac{3\nu}{2} + \frac{3}{4}\; I_0 g\sqrt{\nu} \; .
$$
Then the transition frequency (16), for a transition between the ground
state and an excited mode, becomes, respectively,
$$
\om_{100,0} \simeq 2 - 0.031747 \; g\sqrt{\nu} \; , \qquad
\om_{010,0} \simeq 1 - 0.031747 \; g\sqrt{\nu} \; , \qquad
\om_{001,0} \simeq \nu - 0.015874 \; g\sqrt{\nu} \; .
$$
And the transition amplitudes (45) take the values
$$
\al_{0,100} \simeq \al_{0,010} \simeq \al_{0,001} \simeq 0 \; ,
$$
$$
\al_{100,0} \simeq \al_{010,0} \simeq 0.031747\; g\sqrt{\nu} \; , \qquad
\al_{001,0} \simeq 0.015874\; g\sqrt{\nu} \; .
$$
Hence, in the weak-coupling limit, conditions (46) are always valid.

In the strong-coupling limit $g\nu\ra\infty$, for the control functions
we get
$$
u_0 \simeq \frac{2.282947}{(g\nu)^{2/5}} \; , \qquad
v_0 \simeq \frac{2.282947}{(g\nu)^{2/5}}\; \nu^2 \; ,
$$
$$
u_{100} \simeq \frac{5.823454}{(g\nu)^{2/5}} \; , \qquad
v_{100} \simeq \frac{1.941151}{(g\nu)^{2/5}}\; \nu^2 \; ,
$$
$$
u_{010} \simeq \frac{4.565895}{(g\nu)^{2/5}} \; , \qquad
v_{010} \simeq \frac{2.282947}{(g\nu)^{2/5}}\; \nu^2 \; ,
$$
$$
u_{001} \simeq \frac{2.056114}{(g\nu)^{2/5}} \; , \qquad
v_{001} \simeq \frac{6.168342}{(g\nu)^{2/5}}\; \nu^2 \; ,
$$
which shows that in this limit the effective oscillator frequencies
diminish. For the energies (43), we find
$$
E_0 \simeq 0.547539 \; (g\nu)^{2/5} + 
0.570736\; \frac{2+\nu^2}{(g\nu)^{2/5}} \; , \qquad
E_{100} \simeq 0.643949\; (g\nu)^{2/5} + 
0.485287\; \frac{18+\nu^2}{(g\nu)^{2/5}} \; ,
$$
$$
E_{010} \simeq 0.547539\; (g\nu)^{2/5} + 
0.570736\; \frac{8+\nu^2}{(g\nu)^{2/5}} \; , \qquad
E_{001} \simeq 0.607942\; (g\nu)^{2/5} + 
0.514029\; \frac{2+9\nu^2}{(g\nu)^{2/5}} \; .
$$
Notice that passing from the weak-coupling to strong-coupling limit,
the effect of level crossing happens, since the order of energy levels 
can be changed. Specifics of the level crossing depend on the value 
of the aspect ration $\nu$. For instance, if the trap is cigar-shape 
$(\nu\ll 1)$ then the arrangements of the energy levels in the 
weak-coupling limit,
$$
E_0 < E_{001} < E_{010} < E_{100} \qquad (g\ra 0) \; ,
$$
changes, in the strong-coupling limit, to
$$
E_0 < E_{010} < E_{001} < E_{100} \qquad (g\ra \infty) \; .
$$
The transition frequencies in the strong-coupling limit are
$$
\om_{100,0} \simeq 0.096410\; (g\nu)^{2/5} \; , \qquad
\om_{010,0} \simeq \frac{3.424416}{(g\nu)^{2/5}} \; , \qquad
\om_{001,0} \simeq 0.060403\; (g\nu)^{2/5} \; .
$$
For the integral (44), we have
$$
J_{100} \simeq 0.052069\; , \qquad J_{010} = 0.056439 \; , \qquad
J_{001} \simeq 0.053063 \; .
$$
Then the transition amplitudes (45) become
$$
\al_{0,100} \simeq 0.140202\; (g\nu)^{2/5} \; , \qquad
\al_{100,0} \simeq 0.101636\; (g\nu)^{2/5} \; , 
$$
$$
\al_{0,010} \simeq 0.170345\; (g\nu)^{2/5} \; , \qquad
\al_{010,0} \simeq 0.170345\; (g\nu)^{2/5} \; , 
$$
$$
\al_{0,001} \simeq 0.147059 \; (g\nu)^{2/5} \; , \qquad
\al_{001,0} \simeq 0.122898 \; (g\nu)^{2/5} \; .
$$
This shows that in the asymptotic limit $g\nu\ra\infty$, inequalities 
(46) do not hold. They become valid outside the region of convergence
of this asymptotic limit. The corresponding condition (48), for the
considered lower modes, yields
$$
|g\nu| \leq 1.1\; \left ( 2+\nu^2 \right )^{5/4} \qquad (p=1,\; q=1) \; ,
$$
$$
|g\nu| \leq 0.7\; \left ( 18+\nu^2 \right )^{5/4} \qquad (p=3,\; q=1)\; ,
$$
$$
|g\nu| \leq 1.1\; \left ( 8+\nu^2 \right )^{5/4} \qquad (p=2,\; q=1) \; ,
$$
$$
|g\nu| \leq 0.8\; \left ( 2+9\nu^2 \right )^{5/4} \qquad (p=1,\; q=3) \; .
$$
These inequalities demonstrate that large values of the coupling
parameter $g\gg 1$ can always be compensated by accepting an appropriate
trap shape, being either cigar-shaped $(\nu\ll 1)$ or pancake-shaped
$(\nu\gg 1)$.

\newpage

\begin{center}

{\large{\bf Figure Captions}}

\end{center}

\vskip 1cm

{\bf Fig. 1}. Characteristic evolution of the phase portrait on 
the population difference, $s(t)$, - phase difference, $x(t)$, plane, 
under zero detuning $\ep=0$ and varying amplitude of the resonant field:
(a) $b=0.1$, (b) $b=0.4$; (c) $b=0.49$; (d) $b=0.51$; (e) $b=0.8$.

\vskip 1cm

{\bf Fig. 2}. Dramatic change in the temporal behaviour of the 
population difference $s(t)$ (dashed line) and phase difference $x(t)$
(solid line) when crossing the critical line at the point $\{ b_c=0.5,\;
\ep=0\}$. The initial conditions are $s_0=-1$, $x_0=0$. Time is presented
in dimensionless units, as explained in the text. In Figs. 2(a) and 
2(b), the value of $b$ is, respectively, just below and just above the
critical point $b_c$, with the variation of $\Dlt b=10^{-5}$.

\vskip 1cm

{\bf Fig. 3}. The change of dynamics in the population difference $s(t)$
(dashed line) and phase difference $x(t)$ (solid line) when crossing
the critical line at the point $\{ b_c=-0.5,\; \ep=0\}$. The initial
conditions are $s_0=-1$, $x_0=0$. Time is dimensionless (see the text).
In Figs. 3(a) and 3(b), the value of $b$ is, respectively, slightly above
and slightly below the critical point $b_c$, with the variation
$\Dlt b\sim 10^{-5}$.

\vskip 1cm

{\bf Fig. 4}. The squeezing factor $Q_z$ as a function of time for the
transition amplitude $b=0.49$, zero detuning, and the initial conditions
$s_0=-1$ and $x_0=0$.

\end{document}